\newif\ifdebug\debugfalse
\newif\ifreview\reviewfalse
\ifreview\debugfalse\fi
\documentclass{llncs}
\usepackage[resetfonts]{cmap}
\usepackage[utf8]{inputenc}
\usepackage[T1]{fontenc}
\usepackage{amsmath}
\usepackage{amsfonts}
\usepackage{ifpdf}
\ifpdf
    \usepackage[protrusion,expansion,step=1]{microtype}
\else
    \usepackage[protrusion]{microtype}
\fi
\usepackage{xcolor}
\usepackage{graphicx}
\usepackage{grffile}
\usepackage{subcaption}
\usepackage[normalem]{ulem}

\usepackage[backend=biber,
            style=numeric,
            maxnames=2,
            alldates=iso8601]{biblatex}
\DeclareFieldFormat{labelnumberwidth}{#1\adddot}
\usepackage[unicode=true,
            plainpages=false,
            pdfpagelabels=true,
            colorlinks=false,
            pdfpagemode=UseNone,
            pdfstartview={XYZ null null 1},
            pdfpagelayout=OneColumn,
            pdfdisplaydoctitle=true,
            bookmarks=true,
            bookmarksopen=true,
            bookmarksopenlevel=5,
            bookmarksnumbered=true
           ]{hyperref}

\ifdebug
    \long\def\TODO#1#2{\par\noindent\bgroup{{\bf TO be DOne by \textcolor{red}{#1}:}}\\\it\color{blue}#2\egroup\par}
    \long\def\DONE#1#2{\par\noindent\bgroup{\color{gray}\sout{\bf DONE by \textcolor[rgb]{1,.6,.6}{#1}:}}\\\it\color[rgb]{0.38,0.95,1}{\sout{#2}}\egroup\par}
    \long\def\NOTE#1#2{\par\noindent\bgroup{{\bf NOTE by \textcolor{red}{#1}:}}\\\it\color{green}#2\egroup\par}
\else
    \long\def\TODO#1#2{\relax}
    \long\def\DONE#1#2{\relax}
    \long\def\NOTE#1#2{\relax}
\fi

\newcommand{\orcid}[1]{\href{https://orcid.org/#1}{ORCID~iD:\,#1}}
\renewcommand{\slash}{/\hskip0pt\relax}
\newcommand\Dash{---}
\newcommand*\rot{\rotatebox{70}}
\newcommand{\fone}{\ensuremath{\text{F}_1}}

\newcommand{\var}[1]{\ifmmode\mathit{#1}\else\textit{#1}\fi}
\newcommand{\op}[1]{\ensuremath{\operatorname{#1}}}

\newcommand{\weightt}%
    {\ensuremath{\var{weight}_{\text{token}}}}
\newcommand{\weightm}%
    {\ensuremath{\text{\texttt{mias\_weight}}}}
\newcommand{\dMSCa}%
    {\var{diagonal\-Area}}
\newcommand{\ndMSCa}%
    {\var{nonDiagonal\-MSCArea}}
\newcommand{\SMQ}%
    {\var{Similarity\-Matrix\-Quality}}
\newcommand{\avg}%
    {\op{avg}}
\newcommand{\round}%
    {\op{round}}
\newcommand{\mtmod}%
    {\op{MTermWghtMod}}

\ifreview
    \newcommand\Authors{No Author Given}
\else
    \newcommand\Authors{Michal R{\r u}{\v z}i{\v c}ka; Petr Sojka}
\fi
\newcommand\Title{Towards Math-Aware Automated~Classification and Similarity~Search of Scientific~Publications}
\newcommand{\Subtitle}{Methods of Mathematical~Content~Representations}
\newcommand{\Subject}{CICM 2017 Paper}
\newcommand{\Keywords}{%
  information retrieval;
  representations of mathematics;
  machine learning;
  topic models;
  Gensim;
  TfIdf;
  LSI;
  LDA;
  MSC;
  arXiv.org}

\title{\Title}
\subtitle{\Subtitle}

\ifreview\else
\author{
	Michal R{\r u}{\v z}i{\v c}ka
    \and
	Petr Sojka
}
\institute{
    Faculty of Informatics, Masaryk University,\\
    Botanick\'a 68a, 602\,00 Brno, Czech~Republic\\
    \email{mruzicka@mail.muni.cz},
    \orcid{0000-0001-5547-8720}\\
    and
    \email{sojka@fi.muni.cz},
    \orcid{0000-0002-5768-4007}
}
\fi
\let\outcome\emph % zdurazneni celych vet 

\begin{document}
\pagestyle{plain} % for easy handling and orientation
\hyphenation{lower-cased Infty-Reader meth-odologies wrap-ped} % correct bad hyphenation here
\hypersetup{%
  pdftitle={\Title: \Subtitle},
  pdfauthor={\Authors},
  pdfkeywords={\Keywords},
  pdfsubject={\Subject}}

% Research questions:
% U STEM dokumentů je matematický zápis podstatnou (hlavní) částí sdělení.
% Bude formální zápis matematiky fungovat pro strojové učení klasifikace
%  1) místo textu?
%  2) spolu s textem?
% Odpopvěď:
%  1) Ne, lepší jsou top-MTermy než TeX
%  2) Ano, jen mírně, opět jsou lepší top-MTermy než TeX

\maketitle

\begin{abstract}
In this paper, we investigate mathematical content representations suitable for
the automated classification of and the similarity search in STEM documents
using standard machine learning algorithms: the Latent Dirichlet Allocation
(LDA) and the Latent Semantic Indexing (LSI).

The methods are evaluated on a subset of arXiv.org papers with the Mathematics
Subject Classification (MSC) as a reference classification and using the
standard precision/recall/\fone-measure metrics.

The results give insight how different math representations
may influence performance of the classification and similarity search tasks
in STEM repositories.
Non-surprisingly, machine learning methods are able
to grab distributional semantics from textual tokens.
A proper selection of weighted tokens representing math
may improve the quality of the results slightly.
A structured math representation that imitates successful
text-processing techniques with math is shown to yield better
results than flat \TeX\ tokens.
%The combination of text and thoroughly selected canonicalized math representants 
%achieved the highest score in our experiments. A variety of tested experimental 
%setups provides an overall view of the behavior of different machine learning 
%methods on various data representations.
\end{abstract}

\TODO{PS}{Důležitost a motivace MSC, popis, co je MSC.}
\TODO{PS}{Motivace na příkladech – k čemu to může být dobré:
  na co je automatická klasifikace a podobnostní hledání.}
\DONE{MR}{\emph{Zdůvodnit a rozepsat smysl MTermů.} Přidat vysázené příklady 
  reperentovaných formulí.}
\DONE{PS}{Zrušit kurzívu.}

\section{Introduction}
There are many machine learning techniques usable for the automatic 
classification.~\cite{ml:sebastiani02machine}
Historically, the bag of words (BoW) technique was used with metadata and full texts.
Papers from the Science, Technology, Engineering, and Mathematics (STEM) domain 
are specific in the massive use of mathematical expressions.
However, these parts of documents were usually omitted altogether even though 
they may be highly important for the proper document representation. 
It was shown, for example, that the performance of 
classifiers depends even on the semantic drift of concepts in a language over
time.~\cite{ir:salles2010automatic}
Watt~\cite{dml:watt2008dml} has shown that even a sole histogram of
mathematical symbols that were used in a paper is
sufficient for a rough classification of scientific papers into mathematical
subdomains.

In this paper, we investigate the usefulness of various representations of 
mathematical content (see Section~\ref{sec:math-formats}) for the automated 
classification of and the similarity search in STEM documents
using machine learning algorithms.
We use a human-assigned MSC (see Section~\ref{sec:dataset}) as a reference 
for measuring the quality of the results.
In Section~\ref{sec:algorithms}, this known classification is used as
a reference for evaluating the results of processing various document
representations specified in Section~\ref{sec:preprocessing} with
different setups of the Latent Dirichlet Allocation (LDA) and the
Latent Semantic Indexing (LSI) machine learning algorithms.
To evaluate the results, we use the standard metrics: precision,
recall and the micro \fone-measure in Section~\ref{sec:evaluation}.
Results are discussed in Section~\ref{sec:results} followed by conclusions in 
Section~\ref{sec:conclusions}.

\section{STEM Document Representations}
\label{sec:math-formats}

\DONE{PS}{Add related work.}
\TODO{PS}{Zdůvodnění baselines.}

The transformation of a text document to a BoW is often more complex
than splitting the text to individual word tokens.
It is necessary to cope with punctuation, to `normalize' the words
transforming all characters to lowercase, to apply stemming or lemmatisation, etc.
These text-processing techniques are well established,~\cite{Mitchell1997ml} 
but the processing and the representation of mathematical content have not
yet been properly explored.
For expressing math \TeX\ markup is used on the authoring side, and
MathML~\cite{dml:mathml-3.0} for web document exchange.
When indexing documents with math for search, many representations have
been explored and used~\cite{Zanibbi:2016:MMF:2911451.2911512,ZB11}.
The markup varies from presentation to content/semantic one,
with prevalence availability of presentation notation, available as
a common denominator of conversions from different primary document formats.

We experimented with two different math statement representations:
\begin{description}
    \item[\TeX] Compact linear plain text \TeX\ notation is well known to the
        authors of STEM documents. The vast majority of papers in math is 
        authored in the \TeX\ markup.

        For example the \TeX\ representation of the simple formula
        $a + b^{2 + c}$ is:\newline
        \verb|$a + b^{2 + c}$|

    \item[MTerms] The MTerms are a special compressed and generalized encoding 
        of formulae based on the original MathML 
        representation of math in our data corpus.

        For example, the MTerm representation of the simple formula
        $a + b^{2 + c}$ is:\newline
        \verb|R(I(a)O(+)J(I(b)R(N(2)O(+)I(c))))|
\end{description}

MTerms, described 
in~\cite[p.\,233]{dml:sojkaliska2011}, were developed as an internal math 
formulae representation of the math-aware full text search system 
MIaS.~\cite[p.\,234 bottom]{dml:sojkaliska2011}

An important feature of MIaS MTerms processing is \emph{the canonicalization of 
mathematical content}~\cite{FormanekEtAl:OpenMathUIWiP2012} that is incorporated 
in the MTerms derivation process.
The canonicalization ensures that identical terms 
can be found in distinct documents sharing not just identical, but also
\emph{only similar} subcomponents of otherwise distinct complex formulae.

\begin{figure*}[tb]
    \centerline{\includegraphics[width=\textwidth]{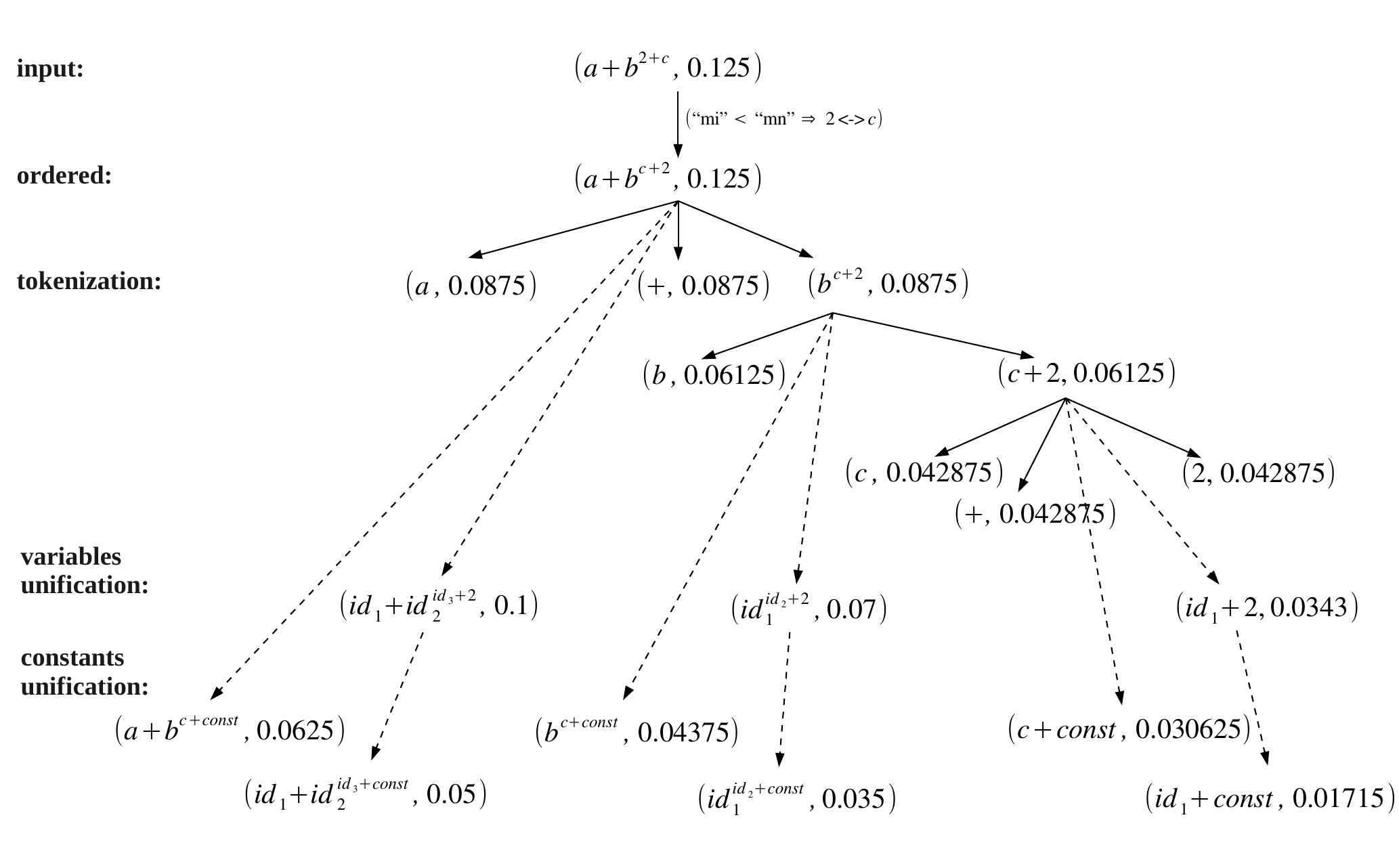}}
    \caption{An example of formula preprocessing inside the MIaS system. During the 
          processing, the order of components in the formula is normalized and 
          subformulae are derived. The last two steps of the processing add 
          generalized forms of all the (sub)formulae with variables and constants 
          unified to generalized identifiers. For more details,
          see~\cite{doi:10.1145:2034691.2034703}. Derived components have 
          a derived MIaS weight assigned, reflecting the degree of change compared to
          the original formula. All these components starting at the ordering 
          level are consequently converted to the MTerm encoding and provided 
          together with the particular MIaS weights as weighted MTerms}
    \label{fig:mias-formula-processing}
\end{figure*}

Another important feature of the MIaS implementation is the ability to 
\emph{derive multiple MTerms from a single input 
formula}.~\cite{doi:10.1145:2034691.2034703}
These additional MTerms represent \emph{subformulae (subcomponents) 
of the original formula} and their \emph{generalized (unified) variants}
with variables and constants substituted with general identifiers.
See Figure~\ref{fig:mias-formula-processing}.

To reflect differences between the \emph{top-MTerms}, i.e.\ MTerms representing 
the original formulae from the dataset (i.e.\ the formula $a+b^{c+2}$ in 
Figure~\ref{fig:mias-formula-processing}), and \emph{derived-MTerms} 
representing the derived forms of these formulae (the formulae at the lower 
levels of the processing tree in Figure~\ref{fig:mias-formula-processing}), the MIaS 
system assigns a $\mathit{MIaS\ weight} \in \langle 0, 1 \rangle 
\subset\mathbb{R}$ to both top- and derived-MTerms. The MIaS weight reflects the 
distance of a particular MTerm
from the original formula. The lower the weight, the greater the distance.%
\footnote{Please note, however, the top-MTerm weight is not 1.0. The initial 
          weight of the formula is computed based on the number of
          subcomponents of the formula. See 
          Figure~\ref{fig:mias-formula-processing}, for more details
          see~\cite{doi:10.1145:2034691.2034703}.}
Weighted MTerms are tuples of the MTerm and its MIaS weight, i.e.:
\begin{verbatim}
{ "mias_weight" : 0.125,
  "mterm"       : "R(I(a)O(+)J(I(b)R(I(c)O(+)N(2))))" }
\end{verbatim}
in the case of the top-MTerm representing formula $a + b^{c + 2}$ in 
Figure~\ref{fig:mias-formula-processing}.

To sum things up, the MTerms representation covers \emph{not only the original formulae 
but also a variety of its derivatives.}

We exploited the existing implementation of the formulae normalization and 
decomposition in the MIaS to imitate some successful
text-processing techniques (such as the normalization of word forms) with formulae.
We tested various strategies for selecting MTerms from the dataset.
MTerms and the \TeX\ statements alone were used as the final tokens without any 
further processing.

\section{Dataset}
\label{sec:dataset}

In our experiments, we used the \href{https://arxiv.org/}{arXiv.org} STEM 
articles with known MSC codes~\cite{www:msc} as a reference classification. 
Their original \TeX\ format 
is hard to fully process without a full-featured \TeX\ compiler,
so we used the NTCIR-11 Math-2 dataset%
\footnote{The dataset is available free of charge for research purposes:
          \url{http://research.nii.ac.jp/ntcir/permission/ntcir-11/perm-en-MATH.html}.}%
~\cite[p.\,89]{NTCIR11-Overview} of 105,120 scientific articles from arXiv.org.
The articles in the dataset were converted to the XML-based XHTML5 
format with mathematical formulae converted to MathML annotated with \TeX.

We selected 8,004 articles from our dataset with exactly one MSC assigned. 
Since they are assigned by the document authors themselves, we assume that the MSC codes are
valid and that they match the document content.
Out of these documents, we selected those with `good' MSC codes in which the third character 
of a code was \emph{neither} \verb|-|~(hyphen) \emph{nor} \verb|.| (period) and 
the MSC specification available~\cite{www:msc} did \emph{not} 
contain any `see also' references to other MSC codes.
The aim of this filtering was to have correct reference data excluding `weakly
defined' codes.
The assignment of exactly one primary MSC code to a document indicates that the document 
content fits well enough to exactly one MSC category to be used as a reference 
representation of the category.

5,619 articles from our dataset met all these criteria and were used as a data 
corpus for our experiments.

\section{Data Preprocessing and Document Representation}
\label{sec:preprocessing} 
All documents were transformed from the XHTML5 format to an internal
representation consisting of:
\begin{description}
    \item[metadata] the basic document metadata such as the unique document ID, the MSC 
        classification, the title, the authors, and the abstract,
    \item[text] the plain-text representation of the full text of the document
        with all math formulae removed, and
    \item[math] the math formulae in the \TeX\ and MTerms formats (see 
        Section~\ref{sec:math-formats}).
\end{description}

All token types (text, \TeX{}, and MTerms) used in a particular 
experimental setup were put to a single common BoW.
Math tokens were prefixed and suffixed with a single \verb|$| character to 
keep even trivial tokens such as a variable~$a$ (resulting in the token~\verb|$a$|) 
distinct from the text word~`a' (resulting in the token~\verb|a|).

Text elements were tokenized using the Unitok tool.~\cite{michelfeit2014text} 
Segmentation to sentences was not performed and instead all tokens across all
sentences were merged into a single flat BoW.

\TODO{?}{Následující odstavec přepsat bez formulky, přidat slovní vysvětlení.}
In some of our experimental setups, we selected only a subset of the available 
MTerms
based of their final weights. The final weight \weightt\ of a particular MTerm 
token was computed as follows:
$$
    \weightt = \left\lceil\round(\mtmod(\weightm))\right\rceil,
$$
where \weightm\ is the MIaS weight of the MTerm and \mtmod() is a function
whose definition varies across the experimental setups.
A particular MTerm was added \weightt\ times to the BoW.

\section{Algorithms}
\label{sec:algorithms}

For our experiments, we used Gensim~\cite{ml:rehureksojkagensim2010}, a
popular state-of-the-art vector space and 
topic modelling toolkit.
In particular, we were comparing the vector space model transformation of BoW
on various representations of the input data using the
Latent Semantic Indexing (LSI)~\cite{ml:deerwester90indexing} and the
Latent Dirichlet Allocation (LDA)~\cite{ml:lda} transformations both with and without the
Term Frequency--Inverse Document Frequency transformation~\cite{ml:tfidf}
(TfIdf-LSI, TfIdf-LDA).
The input data were indexed for cosine measure document similarity queries using the
\href{https://radimrehurek.com/gensim/similarities/docsim.html\#gensim.similarities.docsim.Similarity}%
{\texttt{gensim.similarities.docsim.Similarity}} interface of Gensim.

The input formulae are occasionally very long.
To reduce the dictionary size, math tokens longer than their hex string
representation in MD4 were replaced with their MD4 hashes.

\section{Evaluation Setup}
\label{sec:evaluation}

We exploit the known MSC of the documents in our dataset as 
a reference classification for our evaluation.
An MSC code being assigned to a document 
signifies that the topic (content) of the document is similar to other 
documents with the same MSC code assigned. 
Thus, we expect that a suitable setup of 
the machine learning methods should distribute documents with similar content 
to groups of similar documents that will correspond to the
grouping induced by the MSC classes.

The machine learning process on our dataset results in a number
\hbox{$s \in \langle 0, 1 \rangle \subset\mathbb{R}$} for every pair of 
documents from the testing corpus, expressing the \emph{mutual similarity of the 
two documents}.
The higher the number is, the more similar the two documents are.

\begin{figure*}[tb]
    %\begin{minipage}{0.49\textwidth}
        \centerline{\includegraphics[height=.49\textwidth]{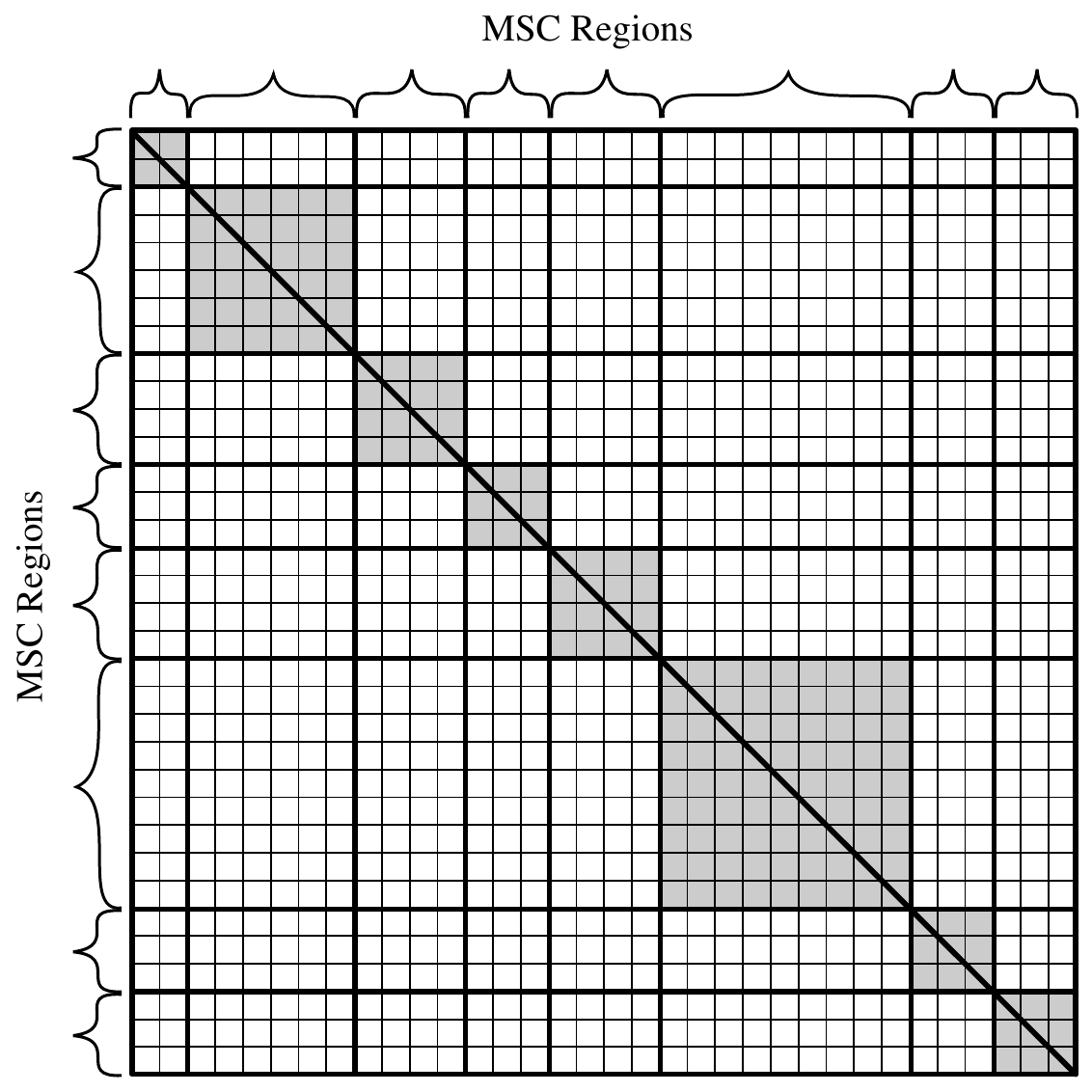}}
        \caption{The similarity matrix schema. The thin grid represents borders 
                 between individual articles, the thick grid represents borders between 
                 MSC regions (groups of articles in different MSC top-level 
                 categories), and gray squares represent the MSC regions
                 grouping the similarity values of articles with identical MSC
                 top-level categories}
        \label{fig:similarity-matrix-schema}
    %\end{minipage}
\end{figure*}

\begin{figure*}[tb]
    \hfill
    \begin{subfigure}[t]{0.5\textwidth}
        \captionsetup{justification=raggedright}
        \centerline{\includegraphics[width=\textwidth]{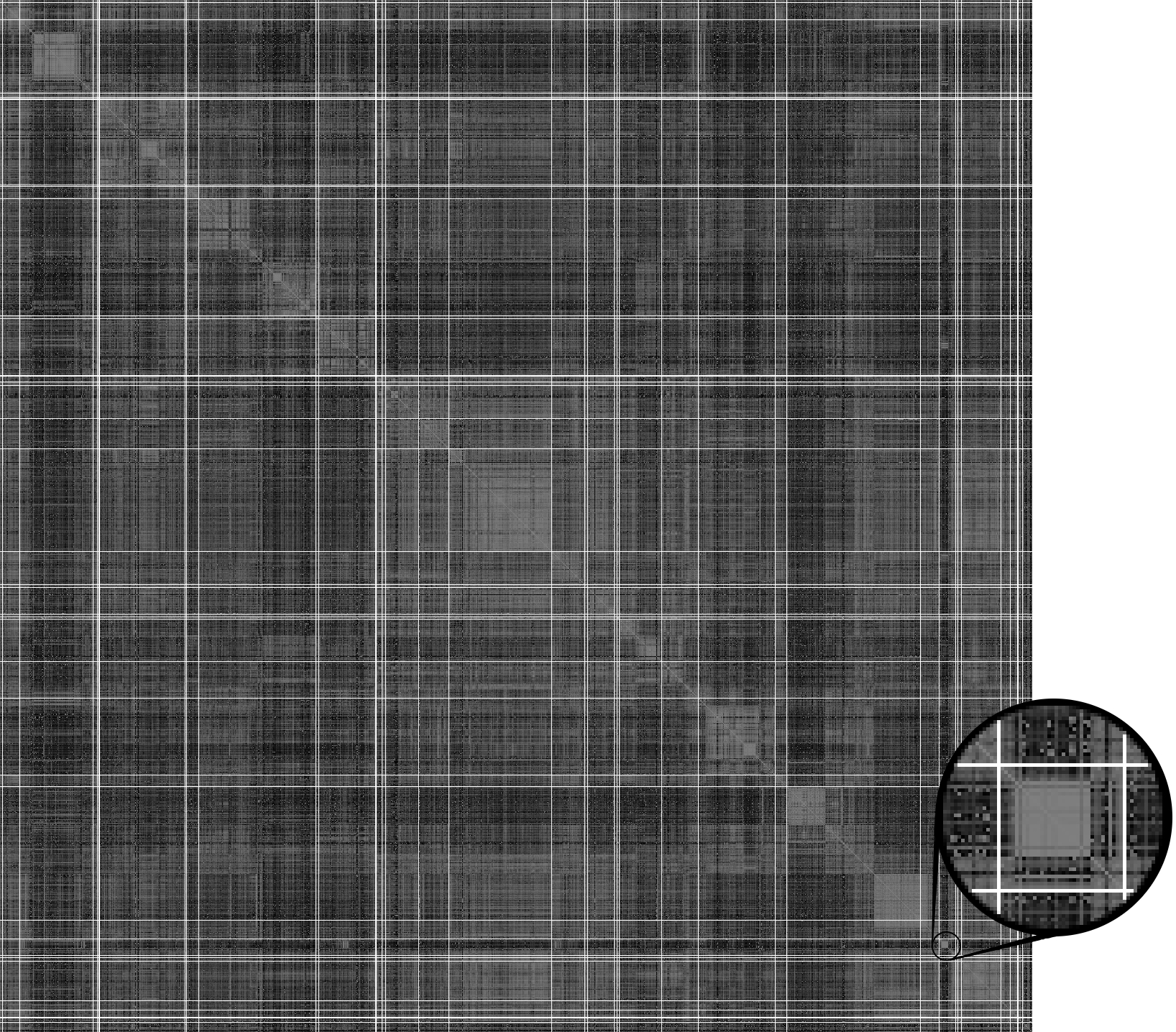}}
        \caption{A successful experimental setup (configuration \#11 in Table~\ref{tab:results:basic}).}
        \label{fig:similarity-matrix}
    \end{subfigure}
    \hfill
    \begin{subfigure}[t]{0.48\textwidth}
        \captionsetup{justification=raggedright}
        \centerline{\includegraphics[width=.92\textwidth]{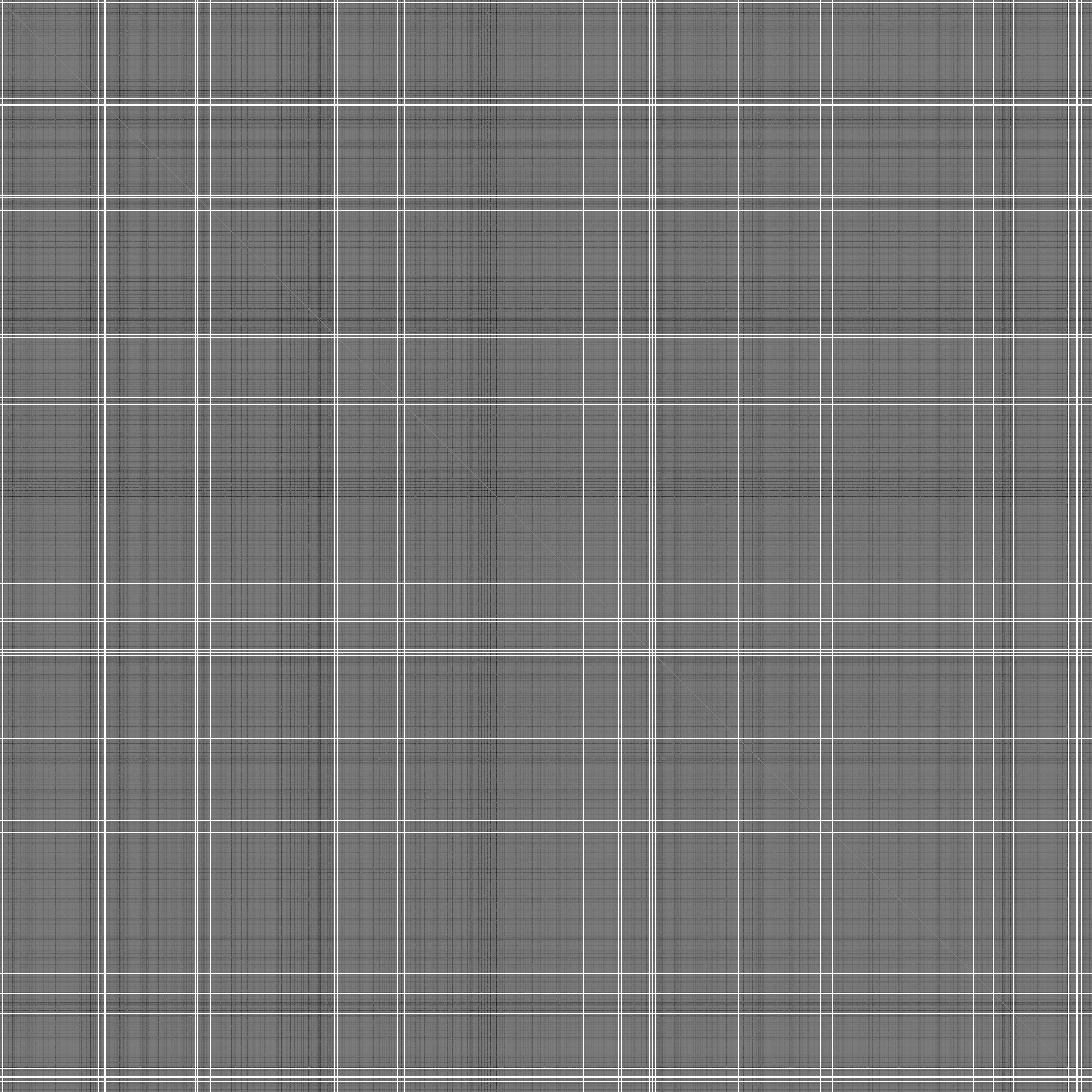}}
        \caption{An unsuccessful experimental setup (configuration\,\#6 in~Table~\ref{tab:results:basic}).}
        \label{fig:similarity-matrix-bad}
    \end{subfigure}
    \hfill
    \caption{A visualization of real similarity matrices from selected results. 
             Bright pixels represents the mutual similarity of a pair of MSC-ordered
             documents at given a row\slash column; the more bright the pixel, 
             the more similar the two documents are. Completely white 
             horizontal/vertical lines represent borders between the MSC
             top-level category regions.
             The magnified circular area in 
             Subfigure~\ref{fig:similarity-matrix} shows a sharply bounded subcategory
             in the region for the top-level MSC code of 68 (Computer Science); this subcategory
             was automatically detected by the machine learning method.}
    \label{fig:similarity-matrices}
\end{figure*}

\begin{figure}[tb]
    \centerline{\includegraphics[width=.8\textwidth]{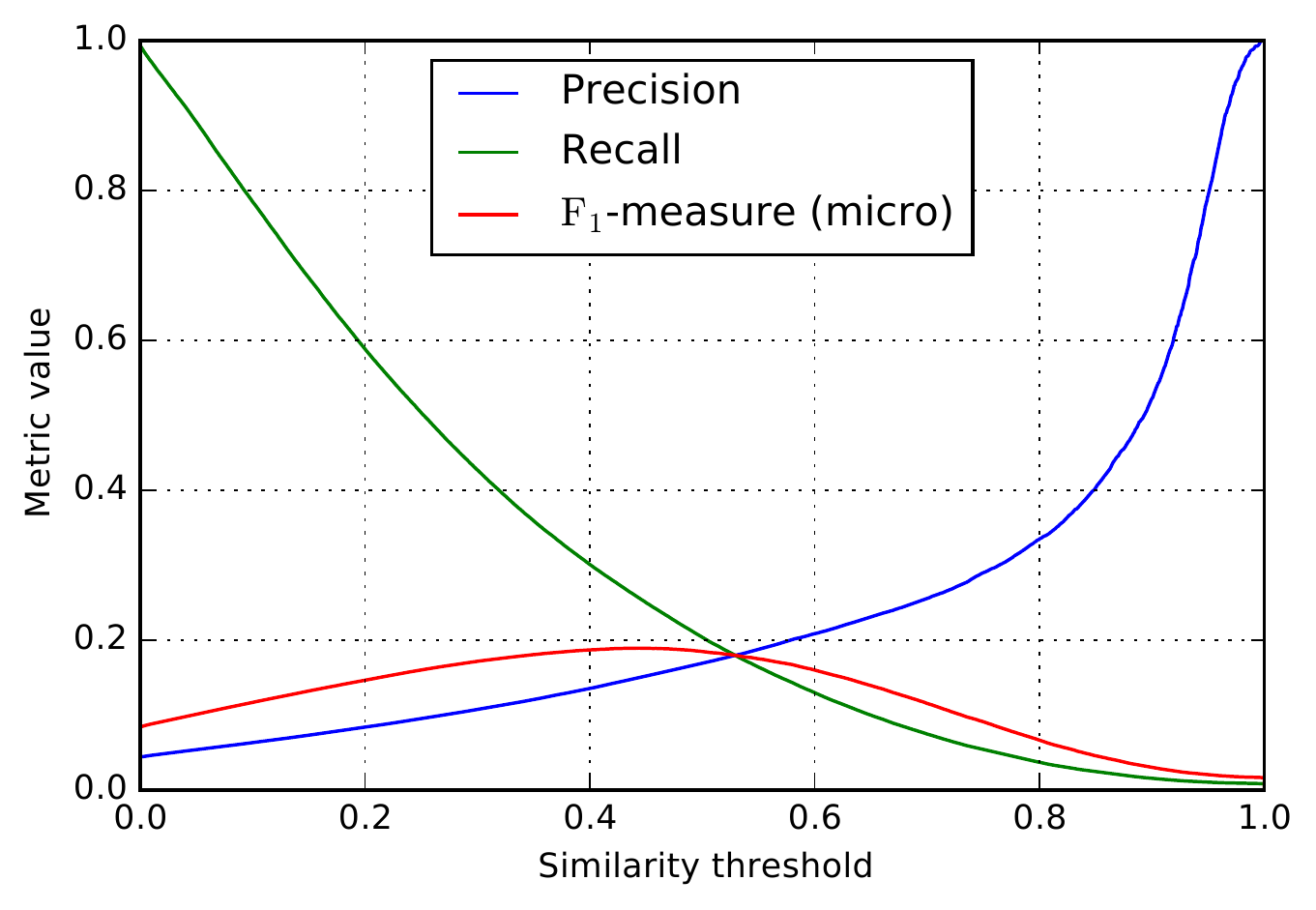}}
    \vspace*{-.9\baselineskip}
    \caption{The precision/recall/\fone-measure curves of one of the tested MTerms 
             content representations (configuration \#16 in 
             Table~\ref{tab:results:basic})}
    \label{fig:metrics}
\end{figure}

\begin{figure}[tb]
    \centerline{\includegraphics[width=.8\textwidth]{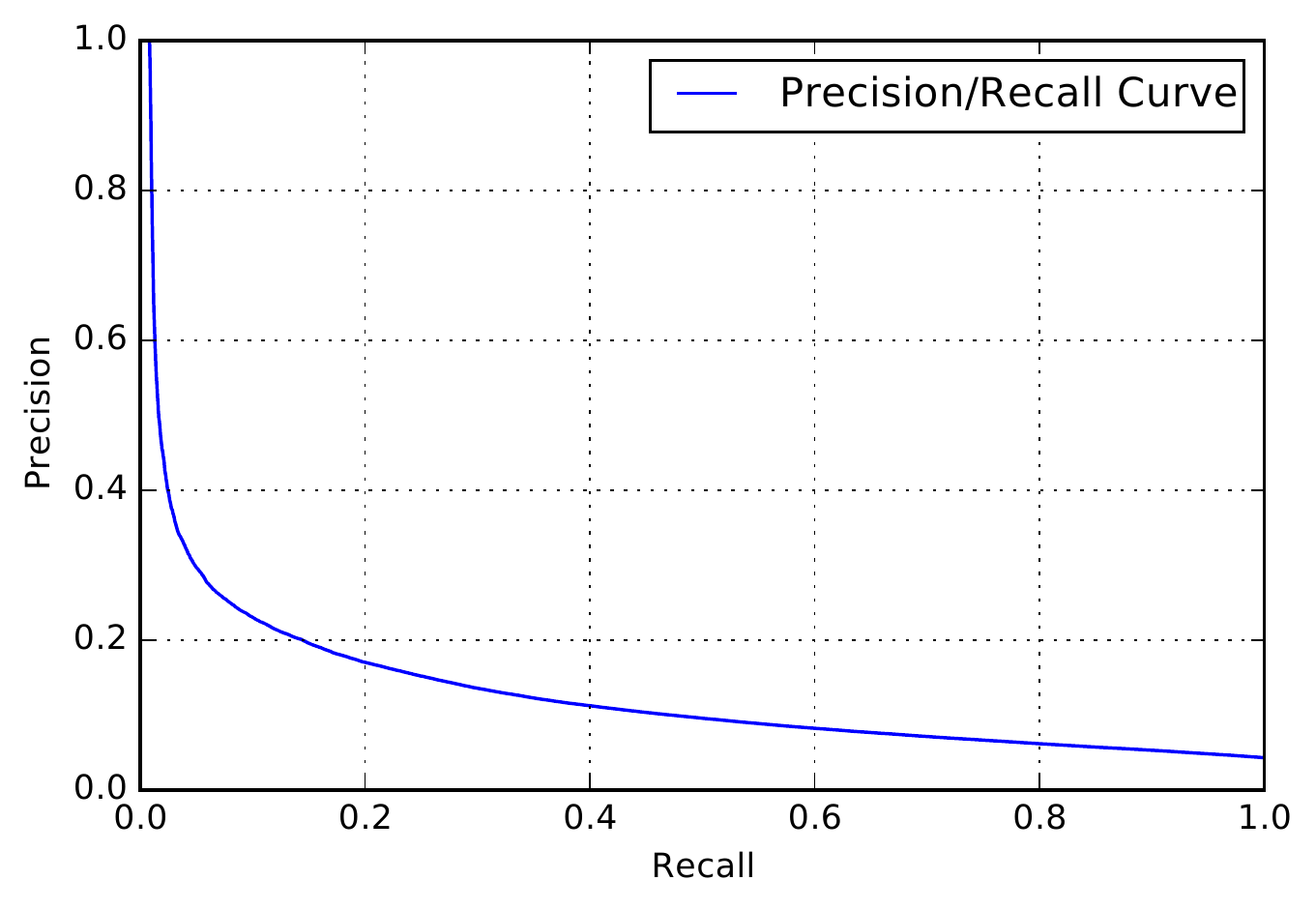}}
    \vspace*{-.9\baselineskip}
    \caption{Precision/recall curves of one of the tested MTerms content 
             representations (configuration \#16 in Table~\ref{tab:results:basic})}
    \label{fig:prec-rec-curve}
\end{figure}

As such, the results can be expressed as a matrix where every column\slash row
belongs to one article from the testing corpus.
Articles in columns\slash rows
of the matrix are ordered according to their MSC code. See
Figure~\ref{fig:similarity-matrix-schema}.
Value $s_{i, j}$, i.e.\ the value in the row~$i$ and the column~$j$ of the
matrix, expresses \emph{the mutual similarity value of the $i$-th and the $j$-th
document according to their MSC ordering}.
As the documents are ordered according to their MSC codes from left to right and 
from top to bottom, the main diagonal of the 
matrix represents the similarity of every document with itself.

We visualized the similarity matrices resulting from our experiments as grayscale images,
transforming the similarity values $s_{i, j}$ to
$\{ 0, \ldots, 255\} \subset \mathbb{Z}$; this value represents the brightness of the pixel in
the images.
To make the image more readable, the transformation was not linear 
but logarithmic to make highly similar pairs of articles more visible as bright 
pixels in the image.
Completely white lines represent borders between MSC top-level category regions, i.e.\ groups of
articles in different MSC top-level categories; these categories are specified by the first two
letters of an MSC code. See Figure~\ref{fig:similarity-matrix}.

\NOTE{MR}{It could be possible useful to add another image showing 
          transformation of Figure~\ref{fig:similarity-matrix} to binary 
          similarity matrix for a given threshold, preferably the one from the 
          results for this configuration.}

The optimal classification method will produce white squares in the MSC regions on 
the main diagonal, indicating high similarity between the articles belonging to 
the same MSC
top-level category, and black rectangles in the MSC regions everywhere else, indicating low
similarity between articles from different MSC top-level categories.
Unsuccessful classification method does not produce these structures in the image:
cf.\ Figure~\ref{fig:similarity-matrix} and \ref{fig:similarity-matrix-bad}.
We will use this observation to rigorously evaluate the quality of results from all 
experimental setups using the standard precision/recall and the
\fone-measure metrics.~\cite{ml:IntroIR2008}

To compute these metrics, we need 
a reference perfect similarity matrix that defines the optimal classification 
result. This matrix has the similarity value of 1 for all items inside the MSC regions on 
the main diagonal and 0 everywhere else.
Figure~\ref{fig:similarity-matrix-schema} shows such a perfect similarity matrix
with the gray areas representing 1s and the white areas representing 0s.

To compute precision/recall and consequently the \fone-measure, we need to select a
threshold $t$ for the transformation of the similarity value $s$ to $s' \in \{1,
0\} \subset\mathbb{Z}$ such that $s' = 1$ if $s \ge t$ and $s' = 0$ otherwise.
In our experiments, we evaluated the results for a large number of samples of 
$t$~values across its possible interval to see the 
precision\slash recall\slash\fone-measure curve shape for different $t$~values.
To compare results of different experimental setups we used results for $t$ at 
precision\slash recall intersection of the particular result.

Using the matrix schema in Figure~\ref{fig:similarity-matrix-schema}, the
gray areas are the relevant items, whereas the white areas are the irrelevant
items.
The values of $s'_{i,j} = 1$ are then true positives inside the gray
areas and false positive inside the white areas; conversely, the values of
$s'_{i,j} = 0$ are true negatives inside the white areas and false negatives
inside the gray areas for the precision\slash recall computation.

Scikit-learn~\cite{scikit-learn} was used to compute precision ($p$) and recall ($r$)
using
\href{http://scikit-learn.org/stable/modules/generated/sklearn.metrics.precision_recall_curve.html\#sklearn.metrics.precision_recall_curve}%
{\texttt{sklearn.metrics.precision\_recall\_curve}}.
Consequently, the \fone-measure in the micro variant is computed at the
given similarity threshold as $ \fone = 2 \frac{p r}{p + r} $.
See the example in Figure~\ref{fig:metrics}.

Prior to the evaluation, we prepared a randomly ordered list of all available documents.
The randomization was done only once at the 
beginning to have a selection that is random but that also stays stable over the course
of all experiments.
Evaluation was done using the $k$-fold cross-validation 
method~\cite{Mitchell1997ml} with $k = 2$, where one fold of the randomly sorted 
documents was used to train the model and the rest was used as a testing corpus.
63 top-level MSC categories were present with an average of 
44~documents per category and a median of 19~documents per category.
To further improve the accuracy of the results (LDA is a probabilistic method)
the $k$-fold cross-validation was run 4~times for every experimental setup.
Consequently, the arithmetic mean and the variance of all values were computed and 
used.

\section{Discussion of Results}
\label{sec:results}

\begin{table}[htbp]
    \renewcommand{\arraystretch}{1.05}
    \tabcolsep2.5dd
    \centering
    
\caption{Experiments uses 50 topics and the 2-fold cross-validation with 4~reruns.
         For the LDA algorithm, gamma\_threshold=0.001, 10~iterations and
           10~passes were used.
%         When weighted MTerms used (`$+$' in MTerms column),
%           the \mtmod() statement %(see Section~\ref{sec:preprocessing}) 
%           was: \mbox{\texttt{trunc(390 * \weightm)}}.
         When top\slash high\slash mid\slash low-MTerms used
           every MTerm was added exactly once to BoW.
         Data are sorted according to the \fone-measure (micro) for
           similarity threshold at precision/recall intersection}
% Výsledky konfigurace:
% 1--99: z runu 2016-05-08-aura
% 100+:  z runu 2016-05-10-mir (reálné èíslo konfigurace je o 100 men¹í)
% 200+:  z runu 2016-05-13-aura-high (reálné èíslo konfigurace je o 200 men¹í)
% 300+:  z runu 2016-05-13-aura-low (reálné èíslo konfigurace je o 300 men¹í)
% 400+:  z runu 2016-05-13-aura-mid (reálné èíslo konfigurace je o 400 men¹í)
\small
\begin{tabular}{|r|c|c|c|l|l|l|l|l|}
    \hline
\rot{Config \#} &
\rot{\TeX} &
\rot{MTerms} &
\rot{Text} &
\rot{Method} &
\rot{$\mathrm{F}_1$ micro at P/R inters. (avg)} &
\rot{$\mathrm{F}_1$ micro at P/R inters. (var)} &
\rot{Maximal $\mathrm{F}_1$ micro (avg)} &
\rot{Maximal $\mathrm{F}_1$ micro (var)} \\
    \hline
115 & $-$ &  top & $+$ & TfIdf-LSI & 0.3427 & 0.0000 & 0.3431 & 0.0000 \\
 \bf 11 & $\boldsymbol{-}$ & $\boldsymbol{-}$ & \bf $\boldsymbol{+}$ & \bf TfIdf-LSI & \bf 0.3419 & \bf 0.0000 & \bf 0.3425 & \bf 0.0000 \\
 15 & $+$ &  $-$ & $+$ & TfIdf-LSI & 0.3366 & 0.0000 & 0.3369 & 0.0000 \\
  8 & $-$ &  $-$ & $+$ & LDA       & 0.3331 & 0.0003 & 0.3356 & 0.0003 \\
  9 & $-$ &  $-$ & $+$ & LSI       & 0.2847 & 0.0000 & 0.2876 & 0.0000 \\
 12 & $+$ &  $-$ & $+$ & LDA       & 0.2523 & 0.0003 & 0.2603 & 0.0002 \\
303 & $-$ &  low & $+$ & TfIdf-LSI & 0.2446 & 0.0000 & 0.2478 & 0.0000 \\
203 & $-$ & high & $+$ & TfIdf-LSI & 0.1922 & 0.0001 & 0.1968 & 0.0001 \\
403 & $-$ &  mid & $+$ & TfIdf-LSI & 0.1913 & 0.0000 & 0.1979 & 0.0000 \\
 24 & $-$ &  $+$ & $+$ & LDA       & 0.1837 & 0.0000 & 0.1922 & 0.0000 \\
 28 & $+$ &  $+$ & $+$ & LDA       & 0.1805 & 0.0001 & 0.1893 & 0.0000 \\
 \bf 16 & $\boldsymbol{-}$ & $\boldsymbol{+}$ & $\boldsymbol{-}$ & \bf LDA       & \bf 0.1743 & \bf 0.0002 & \bf 0.1847 & \bf 0.0001 \\
 20 & $+$ &  $+$ & $-$ & LDA       & 0.1741 & 0.0001 & 0.1839 & 0.0000 \\
111 & $-$ &  top & $-$ & TfIdf-LSI & 0.1568 & 0.0000 & 0.1632 & 0.0000 \\
 27 & $-$ &  $+$ & $+$ & TfIdf-LSI & 0.1535 & 0.0000 & 0.1614 & 0.0000 \\
 19 & $-$ &  $+$ & $-$ & TfIdf-LSI & 0.1533 & 0.0000 & 0.1612 & 0.0000 \\
 23 & $+$ &  $+$ & $-$ & TfIdf-LSI & 0.1532 & 0.0000 & 0.1613 & 0.0000 \\
 31 & $+$ &  $+$ & $+$ & TfIdf-LSI & 0.1528 & 0.0000 & 0.1610 & 0.0000 \\
202 & $-$ & high & $-$ & TfIdf-LSI & 0.1486 & 0.0000 & 0.1562 & 0.0000 \\
 \bf 7 & $\boldsymbol{+}$ & $\boldsymbol{-}$ & $\boldsymbol{-}$ & \bf TfIdf-LSI & \bf 0.1392 & \bf 0.0000 & \bf 0.1512 & \bf 0.0000 \\
  4 & $+$ &  $-$ & $-$ & LDA       & 0.1375 & 0.0001 & 0.1480 & 0.0000 \\
302 & $-$ &  low & $-$ & TfIdf-LSI & 0.1344 & 0.0000 & 0.1453 & 0.0000 \\
402 & $-$ &  mid & $-$ & TfIdf-LSI & 0.1264 & 0.0000 & 0.1395 & 0.0000 \\
  5 & $+$ &  $-$ & $-$ & LSI       & 0.1181 & 0.0000 & 0.1343 & 0.0000 \\
%13 & $+$ &  $-$ & $+$ & LSI       & 0.0968 & 0.0000 & 0.1127 & 0.0000 \\
 10 & $-$ &  $-$ & $+$ & TfIdf-LDA & 0.0888 & 0.0001 & 0.1092 & 0.0001 \\
%14 & $+$ &  $-$ & $+$ & TfIdf-LDA & 0.0784 & 0.0000 & 0.0983 & 0.0000 \\
 22 & $+$ &  $+$ & $-$ & TfIdf-LDA & 0.0664 & 0.0000 & 0.0898 & 0.0000 \\
 18 & $-$ &  $+$ & $-$ & TfIdf-LDA & 0.0657 & 0.0000 & 0.0892 & 0.0000 \\
 26 & $-$ &  $+$ & $+$ & TfIdf-LDA & 0.0655 & 0.0000 & 0.0881 & 0.0000 \\
 29 & $+$ &  $+$ & $+$ & LSI       & 0.0654 & 0.0000 & 0.0910 & 0.0000 \\
%21 & $+$ &  $+$ & $-$ & LSI       & 0.0653 & 0.0000 & 0.0909 & 0.0000 \\
%25 & $-$ &  $+$ & $+$ & LSI       & 0.0653 & 0.0000 & 0.0909 & 0.0000 \\
%17 & $-$ &  $+$ & $-$ & LSI       & 0.0653 & 0.0000 & 0.0909 & 0.0000 \\
 30 & $+$ &  $+$ & $+$ & TfIdf-LDA & 0.0643 & 0.0000 & 0.0875 & 0.0000 \\
  6 & $+$ &  $-$ & $-$ & TfIdf-LDA & 0.0589 & 0.0000 & 0.0850 & 0.0000 \\
    \hline
\end{tabular}

    \label{tab:results:basic}
\end{table}

The basic set of results is shown in Table~\ref{tab:results:basic}.
The table is sorted according to the \fone-measure (micro).
For every setup, the threshold was automatically determined to be the
break-even point at which precision equals recall.
In addition, the maximum \fone-measure (micro) reachable for
each of the results by adjusting the similarity threshold is shown.

2-fold cross-validation with 4~reruns was used to validate the results.
The arithmetic means of the metrics over all the runs are shown in
the `(avg)' columns, a good stability of the method is indicated
by the low variances shown in `(var)' columns.
The best result using text/\TeX/MTerms only are highlighted in bold. 
Configuration \#11 can be considered as the baseline as it represents the 
best result using only text contents of the documents with no involvement of 
mathematical formulae.

\subsection{Properties of the Evaluation Method}
\label{sec:evaluation-method-limits}
We evaluate our results against `the perfect MSC similarity matrix' 
(see Section~\ref{sec:evaluation})\Dash an idealized model that expects all the 
documents belonging to the same top-level MSC category to be perfectly similar 
and completely dissimilar to documents in the remaining top-level MSC categories.
In reality, we cannot expect sharp contrast in the used terminology or even in the
general mathematical notation.

Nevertheless, Figure~\ref{fig:similarity-matrix} 
shows numerous bright sub-regions of highly similar articles inside the
top-level category regions along the main diagonal.
Quite sharply bounded is a sub-square of bright pixels, magnified in the image, 
within the region for the top-level MSC code of 68 (Computer Science).
These indicate the successful determination of MSC 
subcategories showing that the classification method is capable of fine-grained
classification not reflected in our idealized reference model.

Thus, the absolute values of the \fone\ that we reached are of less importance 
for the interpretation of our results.
The relative differences between the various content 
representations are more important for us to determine the usefulness of math 
content for machine learning.

\subsection{Exclusive use of text/\TeX/MTerms}
The best results using only text/\TeX/MTerms are highlighted in bold. 
Configuration \#11 can be considered the \emph{baseline}, as it represents the 
best result using only text content of the documents with no involvement of 
mathematical formulae.

The table shows that the exclusive use of mathematical content
in either \TeX\ or MTerms form cannot supersede the text representation: the \fone-measure at 
the precision\slash recall intersection (hereinafter referred to as \fone) 0.1392 for 
\TeX\ and 0.1743 for MTerms are significantly lower than the 0.3419 for the text.

However, it is important to keep in mind that the tested methods are unsupervised\Dash
we did not provide the system with any hints
on the expected classification. Consequently, the difference in text and 
math content performance may indicate that \emph{(textual) terminology varies more 
between different MSC areas than math notation does and therefore the text content 
is more suitable for the task of determining the top-level MSC category}. 
This interpretation is supported by the observation of a decrease in the \fone\ of the text 
representation after adding unfiltered math content (see the difference between configurations \#11 and \#15 
or between \#8 and \#28 in Table~\ref{tab:results:basic})\Dash 
math, less suitable for this particular task, adds noise, decreasing the quality 
of the classification.

\subsection{Comparison of Machine Learning Methods}

\begin{description}
    \item[TfIdf-LSI] worked best with the combination of text and top-MTerms 
        representations (\#115 in Table~\ref{tab:results:basic}).

    \item[LDA] follows closely with the text only representation.

    \item[LSI] follows the LDA setup with a decrease of 15 \% in terms of \fone with
               the text only representation.

    \item[TfIdf-LDA] yielded poor results in general; $\text{\fone} 
            < 0.1$ in the average case.
\end{description}

% The overall quality of the results could be improved if applied on larger 
% dataset. Comparing 2-fold and 4-fold validation on otherwise identical setups 
% does not change the order of the results but the absolute value of the 
% \fone\ is better roughly by 0.03 in case of 2-fold validated results. We 
% believe that this is due to the higher number of articles available in the 
% training fold in case of the 2-fold setup, resulting in a more accurate model.

For LDA, tuning the gamma threshold (values 0.1, 0.01, and 0.001) and the number 
of iterations (values 10 and 100) had a minimal impact on the results.
In terms of \fone, the difference was less than 0.006 for each parameter with otherwise 
identical setups.

TfIdf-LSI has shown the best performance; we will therefore
concentrate on this method in the next section.

\subsection{Impact of the Number of Topics}
To test the hypothesis that the worse overall performance of math over text content 
is caused by the low number of topics (50 topics were used in most 
setups) limiting the rich structure of math formulae to a small subset, we 
used a batch of configurations with 100, 500, 1,000 and 5,000 topics.

Using TfIdf-LSI, the higher number of topics led to a worse performance using 
text only: \fone\ decreased from 0.3340 for 100 topics to 0.3021 for 5,000 
topics. For MTerms or a combination of MTerms with text, the difference in 
\fone\ was negligible (0.01), although marginally better results were achieved
with 500 topics.

Thus, the low number of topics does not seem to limit performance using math 
representations.

% The difference of up to 0.01 was also measured with TfIdf-LDA for a combination 
% of MTerms and text. However, TfIdF-LDA for text only benefited from the 
% increased number of topics: \fone\
% increased from 0.0960 for 100 topics to 0.1460 for 5,000 topics. That is still 
% a much lower score than only text with TfIdf-LSI. The influence of 
% the number of topics to MTerms was negligible as well.

\subsection{Selection of MTerms}
An important feature the MTerms mathematical content representation has, unlike 
the \TeX\ format, is the normalization of formulae and the derivation of 
subformulae (see Section~\ref{sec:math-formats}): A single input formula is 
converted to a set of weighted MTerms and we may choose to select only a subset 
of the MTerms for the machine learning procedure.

In most experimental setups, we used MTerms filtered according to their weights 
modified with the following definition of \mtmod() in Python:
\mbox{\texttt{trunc(390 * \weightm)}} (see Section~\ref{sec:preprocessing}). 
This definition is the result of our initial analysis of the MIaS subformulae 
weighting~\cite{doi:10.1145:2034691.2034703} and experimental runs aimed to 
include only the original formulae and their first-level derivatives. The final 
weight \weightt\ of a particular MTerm token is reflected in the BoW by adding 
the MTerm token \weightt-times.

We also tested an alternative method of MTerm selection with the TfIdf-LSI approach. 
Every MTerm was considered for inclusion in BoW according to the MIaS weight.
If the MTerm was selected for inclusion, then it was added to the BoW
exactly once regardless of its MIaS weight.

We tested multiple selection strategies:
\begin{description}
    \item[Top-MTerms] This strategy selects only top-MTerms, i.e.\ MTerms 
        representing the original formulae from the dataset after ordering the
        canonicalizing the elements (see Section~\ref{sec:math-formats}).

    \item[High-MTerms] This strategy selects only one third of the 
        highest-weighted MTerms out of the formula set.

    \item[Mid-MTerms] This strategy selects only one third of the middle-weighted 
        MTerms out of the formula set.

    \item[Low-MTerms] This strategy selects only one third of the lowest-weighted 
        MTerms.
\end{description}

% This method reached \fone\ of 0.1533, performing worse than the full
% weighting scheme described in the previous paragraph:
% Selecting only one third of 
% the highest-weighted MTerms out of the formula set reduced \fone\ to 
% 0.1486. Selecting one third of the lowest-weighted MTerms reduced \fone\ even further
% to 0.1344. Selecting only one third of the medium-weighted MTerms led to the
% lowest \fone\ value of 0.1264.

Similarly to the full MTerms set, the selected MTerm strategies reached overall 
higher scores when combined with texts.

However, the exclusive use of top-MTerms (\#111 in Table~\ref{tab:results:basic}) 
reaches \fone\ of 0.1568, i.e.\ performs better than exclusive use the full set of 
the weighted MTerms (\#19 that reached \fone\ of 0.1533). The exclusive use of 
high-/low-/mid-MTerms (\#202, \#302, \#402) follows with the \fone\ scores of
0.1486/0.1344/0.1264. The fact that middle-weighted MTerms score lower than both 
high- and low-MTerms might indicate that the MSC category is better 
determined using full-length formulae (high-MTerms) or the extracted 
basic elements of formulae such as specific symbols (low-MTerms) 
than using the common notation `in the middle' (mid-MTerms).

Using a combination of high-/low-/mid-MTerms with text (\#203, \#303, \#403) 
confirms mid-MTerms to be less useful (reaching \fone\ of 0.1913). However,
low-MTerms reaching \fone\ of 0.2446 work clearly better than high-MTerms reaching 
\fone\ of 0.1922 in this case.

The most interesting setup is the combination of top-MTerms with the text. This 
setup (\#115 in Table~\ref{tab:results:basic}) reached \fone\ of 0.3427, the 
highest out of all of our experimental setups, yielding marginally better results than 
both the text only configuration \#11 (\fone\ 0.3419) and combination of text 
and \TeX (\#15, \fone\ 0.3366).

\section{Conclusions}
\label{sec:conclusions}
The aim of this paper was to investigate mathematical content 
representations for machine learning algorithms. These representations were chosen to be suitable for the automated 
classification of and the similarity search in STEM documents. We tested on a dataset of 
arXiv.org papers with a known MSC. This classification was 
consequently used as the reference classification for the evaluation of various 
setups of our unsupervised machine learning procedure.

For mathematical content representation, we used two distinct formats: 
unstructured \TeX\ representation and the normalized and generalized weighted 
complex structural representation of MTerms.

For our task of an unsupervised classification of STEM papers according to the 
MSC, the text content representation with top-MTerms works best (\fone\ of 
0.3427) but with only very small margin to the exclusive use of text (\fone\ of 
0.3419) or the combination of text and \TeX\ (\fone\ of 0.3366).

Exclusive use of text (\fone\ 0.3419) works better than exclusive use of MTerms 
(\fone\ of 0.1744) or \TeX\ (\fone\ of 0.1392).
However, this may only indicate that \outcome{textual terminology varies more between different MSC areas 
than math notation and as such, text content is more suitable for
the determination of MSC.}

%Even though text representation works better in this particular task 
%mathematical content turned out to be well usable for machine learning of 
%similarity of STEM papers.
%
%In this case, \outcome{simple unstructured \TeX\ representation is less useful than complex 
%MTerms representation benefiting from involvement of the normalization of the 
%original formulae and derivation of properly weighted subformulae.
%
%This effectively imitates successful strategies of text processing
%techniques, such as normalization of word forms, on mathematical contents.}

Even though we did not reach a clear improvement with the use of math content,
\emph{the combination of text and math still achieved the highest score in our 
experiments}. According to our results, this is due to the proper selection of math 
representants~-- the canonicalized top-MTerms.

This leads us to believe that mathematical content has its use in
the machine learning of the similarity of STEM papers. However, to have 
a positive impact, the proper selection of a small number of math content 
representants is needed. We believe that thoroughly selected math tokens added to the 
text tokens can refine the final similarity score in the corner cases where 
the limits of distinguishing based on text content alone are reached. For example, in the case 
of highly formal mathematical articles, the high percentage of the overall 
content are formulae and text only act as a `glue'. Because of that, there are then
few specific text keywords and the importance of the formulae for assigning
the correct similarity score is high.

However, as discussed in Section~\ref{sec:evaluation-method-limits}, we
evaluate against an idealized model on top-level MSC categories. We assume
\outcome{the positive effect of properly selected math content representants on
the corner cases can partially be hidden due to this idealized model}.
We believe finer granularity of the evaluation could uncover clearer positive
effect of our use of math in the process.

Besides searching for better evaluation strategies, our
further research will investigate learning-to-rank methods to pick up the most
representative MTerms.
We will test this method with other features of 
structured data, such as syntactic or dependency trees.
It might be useful to further experiment with the `tokenization' of math to 
smaller semantic elements, possibly better reflecting the specific notation used 
in different disciplines, to effectively imitate successful text processing 
techniques (such as the normalization of word forms) with mathematical content.

{%\raggedright\AtNextBibliography{\small}%
\printbibliography[heading=bibintoc]}
\end{document}